# Interface Transformation from Ruling to Obedience

*Abstract*.  This article is about one feature which was partly introduced 30 years ago with the development of multi windows operating systems.  It is about the movability of screen objects not according to some predetermined algorithm but by the direct user action.  Many years ago it was introduced on a very limited basis and nothing was improved since then.  Smartphones and tablets give direct access to screen elements but on a very limited set of commands (scroll and zoom).  There is an easy to use algorithm which turns any screen object into movable / resizable.  This algorithm uses only mouse to turn screens of normal PCs into touchscreens, but this simple change means a revolution in our work with computers.

Three things can spark a revolution in our dealing with computers.

- New hardware which opens absolutely new possibilities.
- Drastic change of operating system which triggers the redesign of all related applications.
- Some BIG idea which can be implemented inside the existing environment.

The best example of the first case is the mouse invention.

The best example of the second case is the switch from DOS to Windows.  (Apple's Lisa was the first to demonstrate this and related changes, but nearly immediately the same ideas were demonstrated by the Windows system.)  Up till that moment computers were predominantly used as very powerful calculators, so the only requirement for interface was to provide input of some initial data and the output of results.  According to those demands, interface was not important at all and played only a secondary role.  The questions of interface were so unimportant that you can hardly find any articles on the theme written at that period.  At the same time, there was a lot of work on new languages and on using some statements in those languages.  The most valuable results were described in the well known articles [1] and books; the best programming languages were designed at that time.  As for the interface, there was nothing to discuss at all and users, whose number was very limited, had to work with whatever they were given.  Absolute developers' control over interface became an axiom.

Switch to multi windows operating systems changed everything.  Movable icons, movable and resizable windows, opening menus, a set of 10 – 15 widely used controls…  There were so many new opportunities that everything else was moved aside and for many years the interface design became the main theme of research and publications.  The huge number of new users could not work inside the old model.  Different requirements from different groups of users required something new and *adaptive interface* was born.  Innumerable results were reflected in hundreds (more likely thousands) of articles and books, but we have to understand that in multi windows systems there are two non-overlapping areas with absolutely different rules for man – machine interface: the level of operating system and the inner world of each application.

At the level of operating system, there are objects of only two types: small non-resizable rectangular icons and resizable rectangles representing applications.  Non-resizable icons can be moved by any inner point.  Big rectangles can be moved by special area (title bar) and can be resized by any border point.  Users can do this moving and resizing at any moment and in any way they want, so the sizes and placement of all elements at this level are entirely under users' control.  I want to underline once more this very important thing: <u>at the level of operating system moving and resizing of all elements is under full users' control</u>.

The inner world of applications is populated with elements of two different origins: there is a very limited set of controls which are mostly designed by the same developer of operating system and there is an unlimited set of graphical objects born by the imagination and skills of programmers around the world.  Controls are similar in behaviour to the elements from the level of operating system.  All controls are rectangular, some of them are non-resizable; others are resizable; all of them are movable.  The important thing is that controls are movable and resizable by design, but these features are controlled by developers.  It is an extremely rare situation when designers allow users to change the size and position of controls directly.

Graphical objects are developed by programmers all round the world according to the purposes of applications in which these objects are used.  It is an extremely rare situation when such objects are movable.  The best known example of application with movable graphical objects is the *Paint* program.  There are some other programs of the same type, but these are the programs in which the movability of objects is the mandatory requirement.  In each case it requires the work of not simply good but very skilful programmers and the results are applicable only in this particular program.  If you try to remember all PC programs in which you could really move graphical objects, I think that fingers of one hand would be more than enough.

Controls contain the possibility of movability but it is extremely difficult to design movable graphical objects.  Usually the inner world of application is populated by combination of controls and graphical objects.  If only some of the screen elements are movable and others are fixed, then you get a real mess while trying any movement. For this reason the movability of controls is nearly never used and it is a <u>standard situation that inside application nothing is movable</u>.

I purposely underlined two statements. On PCs we have two levels with absolutely different behaviour of elements. It was introduced in such a way 30 years ago and nothing changed since then.

Or maybe things became even worse? There is a huge problem of using the same application on computers with different screen sizes with different resolutions and different fonts. Windows are resizable, but when the inner elements are not movable and not resizable, then the resizing of the whole space does not help at all. There are two solutions to this problem: either the full control over all elements and the whole view is given to users or to developers. The first solution is not easy at all. Very few programmers can develop resizable and movable graphical objects. Development of easy to use algorithm which any programmer can use and apply to an arbitrary graphical object is even much harder. Developers of operating systems could not propose such algorithm years ago and never did it since then. I have no doubts that they thought about it decades ago but could not find a good solution. The difference in movability of elements at two levels was obvious from the beginning. Those were extremely clever people who designed Windows and similar systems. The movability of the screen elements is a very needed thing; if it was not demonstrated as a feature which any programmer could add to any designed element, then those authors of operating systems simply could not find the way to introduce movability as an applicable feature.

If the movability and resizability of arbitrary elements could not be demonstrated but the variations of sizes, resolutions, and fonts caused a big problem in using very well designed programs, then another solution was announced. Around 10 years ago *dynamic layout* was declared as the trend for the future [2]. I would not call it good or bad decision. From my point of view, it is an absolutely wrong decision. Users are stripped of any chances to change the view and have to work with whatever they are provided. They can like or dislike the view of an application but they have to work with the view which developer prefers. That is what we have in the world of PC programs. Nearly all screen elements were introduced 30 years ago and their view and behaviour didn't change since then.

There was a hundred years sleep in one famous fairy tale; we have a comparable time lapse. Meanwhile the life around sleeping castle (PC programs) is going on. Smartphones started by copying the PC interface but quickly went away. These and other small devices give users direct access to every element. The instrument of direct access – a finger – is not too precise and the system of available commands is very limited, but nobody even try to declare that users are too stupid for such direct action with the screen elements and must be banned from it. The idea is very simple: more skilful users can do more things; everyone is working according to his skills and everyone is involved in direct action with the elements.

It is time for PC programs to do the same and to turn all inner elements into movable. Transformation of ALL elements into movable is the third way of interface change which I mentioned at the beginning. Absolutely all elements can be turned into movable / resizable and the full control over those elements (and in this way the full control over applications) is given to users. Mouse is the only instrument which provides such control, so this interface revolution is organized without any change in hardware. Such change of programs will be for the huge benefit of all users. Even more: the advantages are greater for the most complicated applications which are used by the most skilful users to solve the most complex problems. I designed several programs of the new type for researchers from the Department of Mathematical Modelling and they immediately estimated the advantages of such programs. The used algorithm is described in the most detailed way in the book which is accompanied by a big Demo program with all available codes [3].

Any algorithm of such type can become useful only if it is applicable to all elements of arbitrary shape and in all possible situations. To demonstrate such possibilities, the mentioned Demo program includes a huge number of examples. Several figures from the book can illustrate the diversity of using the same algorithm.

New programs are entirely controlled by users, so I call them *user-driven applications*. Such applications have few rules of which the first one is the most important: <u>all elements are movable</u>! Graphical elements are moved by any inner point and resized by borders. Pressing of any control causes some well known reaction and I don't want to interfere with it, so controls are moved and resized by borders.

**Figure 1** includes graphical objects of different shapes. Each element can be moved by inner points and resized by borders. Sector partitions in multicolored circles and rings are movable. Several elements allow reconfiguring by obvious special points (by vertices). There are elements with straight and curved borders; there are solid objects and elements with holes; movability of all of them is provided by the same algorithm.

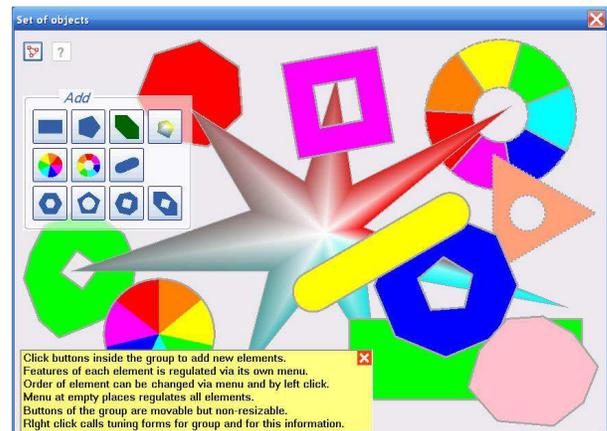

**Fig.1** This example works with different graphical objects, solitary controls, and group of the new type

Movability of elements caused absolutely new ideas in group design. Elements inside the demonstrated group are movable and resizable, so they can be positioned in an arbitrary way. At the same time the frame adjusts its size and position to all changes of inner elements and the whole group can be moved by any inner point.

*Function Analyser* (**figure 2**) allows to analyse Y(x) functions and parametric functions {X(p), Y(p)}. There can be an arbitrary number of plotting areas; each area is movable and resizable. Each area is associated with movable scales; areas and scales can be associated with an arbitrary number of movable and rotatable comments. In the programs with fixed elements, the good positioning of comments along the unknown graphs is a problem without good solution. In the new applications this problem simply does not exist because any comment can be moved to any place and rotated on any angle.

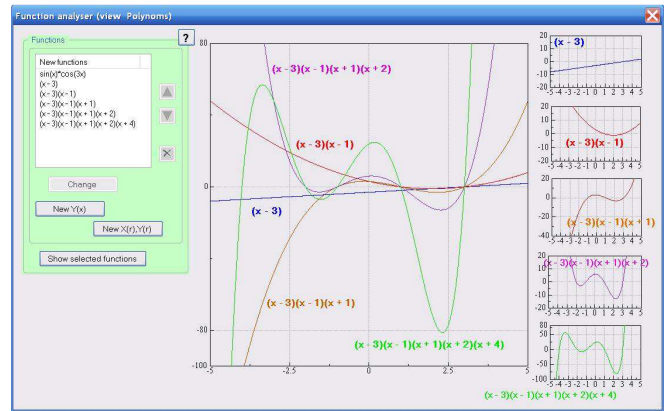

**Fig.2** *Function Analyser*

Engineering and scientific programs were of highest interest for me throughout the whole career while my look at some "financial" plotting (**figure 3**) was only an attempt to find some problematic examples for my algorithm. Unfortunately (or fortunately?) there were no problems at all. Everything is movable, resizable, and rotatable. Users can do with these (and similar) elements whatever they want and introduce the data in any way they prefer. My understanding is that if you work on analysis of some data, then the presentation of this data in the way you personally prefer would be very helpful. Especially if any transformation is done in no time with one or several mouse movements.

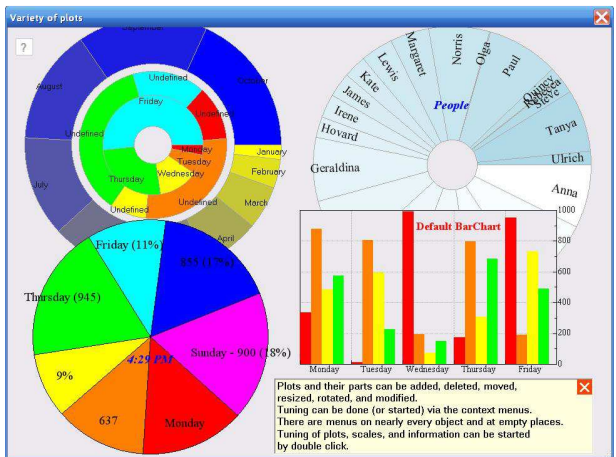

**Fig.3** Plots variety

Everyone knows the view of *Calculator*, so there is definitely something strange in **figure 4**. But the result of pressing some button does not depend on the control position and size, so if I prefer this view and somebody else prefers absolutely different view, the results of calculation have to be the same. In such way each user can organize the view he prefers and change it at any moment, while the designer only guarantees the correctness of results and provides an instrument for easy and quick view change.

Two more examples (**figure 5**) demonstrate that the same algorithm is used in absolutely different areas and with elements which have nothing in common. A small spot can be moved inside labyrinth or along the way. Both objects –

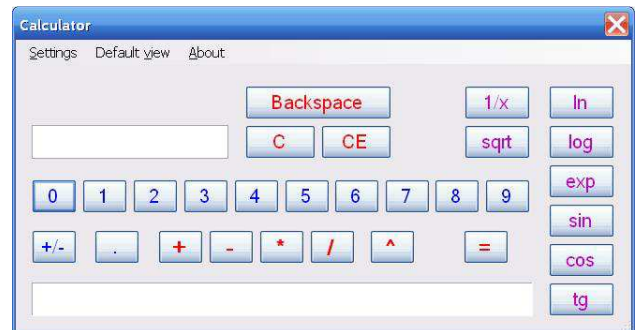

**Fig.4** *Calculator*

labyrinth and the way – can be changed on a fly.

I try to minimize the number of figures in this article but at the same time I want to show that this algorithm can be used with different elements in absolutely different programs. The main thing is that there is no need to design new algorithm for each new example. It is like differentiation and integration which do not depend on the area of their use.

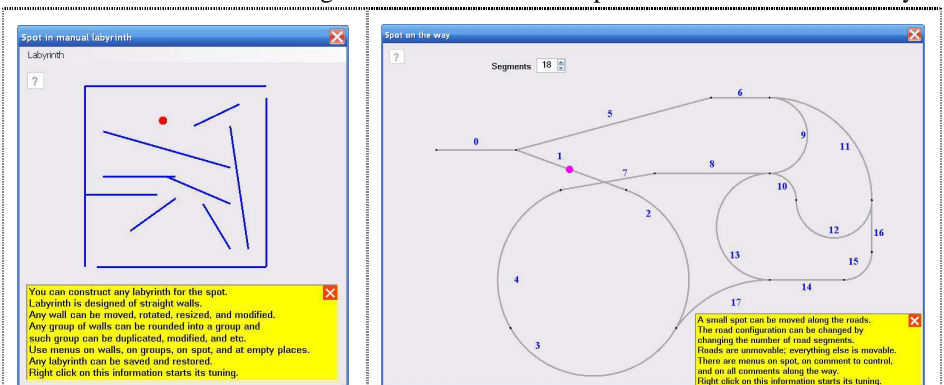

**Fig.5** Small spot can be moved either through labyrinth or along the way

I demonstrate with the book a lot of different examples, but I can't predict which of them can spark greater interest. At one moment I prepared a small *Family Tree* application. The task is well known, but all previous implementations have strict enough limitations on the allowed construction. At the same time, this is just an example of a task with high enough probability of uncertainty about different pieces. This uncertainty requires the construction of the trees which are wrong from the point of Biology but have to be designed in such a way because of the lack of information. My variant of *Family Tree* has only the most obvious limitation (you can't be the parent of yourself), but otherwise users are free to organize any set of people and connections between them. Maybe such flexibility and easiness of design was the cause of very high interest to my *Family Tree*. The same ideas of elements with easily changeable connections can be applied to absolutely different task and the ideas from *Family Tree* were used to design an application for teaching the course on electric circuits.

The proposed algorithm is not aimed at extremely qualified developers. One of amateur programmers took the examples from the very beginning of the book (forward movement of graphical primitives and rotation) and prepared an application for furniture placement in the restaurant he owns.

We constantly move real objects around us. Fixed screen objects in all the programs is only a result of absence of their easy movability. When all screen objects become movable, this feature is constantly used to organize the best view for each task at each particular moment. It is not a burden but absolutely natural use of very helpful feature; exactly as it happens in our everyday life. At the same time, movability of the screen objects is extremely valuable for the most complicated programs in which the data to be shown cannot be predicted because it is the result of some research or calculations. With the unknown input data, only such movability allows to organize the best view for the analysis of the received results.

Now I want to return to the title of this article. From the very beginning of programming history and up till now interface was used to enforce users to work inside the scenario hard coded by developers. At the Golden Age of adaptive interface this awkward situation was camouflaged by giving users a set of choices. The allowed selection softened the problem but didn't eliminate it. Users are still allowed to work only inside the scenario previously approved by developers. For the majority of programs it is the constant source of tensions between developers and users. For the most sophisticated programs which are developed for all branches of science and engineering such situation is simply a disaster because much better specialists in each area (researchers) have to work inside the problem understanding enforced by lesser specialists (program developers). Through the fixed interface, developers have absolute control over the use of applications. Whether you understand it or not, but this is the axiom for all programs with the fixed interface (even if there are several variants).

This axiom is based on the whole history of programs design, but this is absolutely wrong situation. Developers have to provide correct calculations and correct visualization according to users' demands, but this must be the border of developers' control. Users have to get the full control over applications and only users have to decide WHAT, WHEN, and HOW to show. The question of control over applications is simply another formulation of more general question: "What is the main goal of programs development?"

- If developers' income is the main goal, then it is absolutely natural and expected that developers keep control over applications. Under the constant demands for changes from other side such control has to be changed from time to time, so the new forms of control have to be marketed. Adaptive interface, dynamic layout, attempts to add Kinect to every computer, accent on interpretation of eyeball movement or gestures. In each case users are allowed to do something, but the interpretation and transformation of this interpretation into real action is still on developers' side. Developers are ruling through interface.

- If the goal of programs is to fulfil users' tasks, then there must be implementation of direct users' actions without any developer's interpretation. Interface has to provide the ways to get users' commands and then transmit those commands. Nothing else.

I want to illustrate the last statement with the mentioned examples.

In the *Function Analyser* (**figure 2**), user declares the functions and then regulates their demonstration. Only user decides about the number of plotting areas; about positions, sizes, and parameters of these areas; about the set of graphs to be shown in each area. Any set of plotting areas can be saved as a view. User can organize any number of views and then switch between them at any moment.

In the *Calculator* (**figure 4**), user constructs any view he wants; the result of calculations is independent of the current view.

In the *Spot in Labyrinth* (**figure 5** on the left), user can construct any labyrinth he wants; the spot is moved in exactly the same way through an arbitrary labyrinth. Any number of different labyrinths can be designed, saved, and used.

All those mentioned results are achieved by making all the screen objects movable and by giving users full control over this movability. There is no need for new hardware. There is no new operating system and all applications have a familiar view. There is only one new idea. ALL objects become movable and users get direct control over all elements. With this we find ourselves in absolutely new realm of man-machine relation in which we discover a lot of benefits for us – USERS.